%Paper: hep-th/9404186
%From: sethi@string.harvard.edu (Savdeep Sethi)
%Date: Fri, 29 Apr 1994 16:36:27 -0400

%%   This paper requires harvmac.tex and was written     %%
%%   using InstantTeX.                                                      %%
%%   Use the "little" option and save paper!                      %%

\input harvmac.tex

\font\ninerm=cmr9
\def\kh{K\"{a}hler }
\def\lg{Landau-Ginzburg }
\def\c{\hat{c}}
\def\gh{\eta}
\def\sp{{\bf SP}}
\def\met{K_{\alpha{\bar{\beta}}}}
\def\invmet{K^{{\bar{\beta}}\alpha}}
\def\wsp{{\bf WSP}}
\def\sdet{{\rm sdet}}
\def\ch{{\cal R} }
\def\sm{{\cal M} }
\def\b{\hat{c}_b}
\def\cy{{\it  X} }
\def\emb{{\it Y}}
\def\res{{\rm Res }}

%% List of References

\lref\indexthm{S. Sethi, in preparation.}
\lref\cvessays{C. Vafa, in {\it Essays on Mirror Manifolds}, ed. by S-T
Yau, (International Press, 1992).}
\lref\quantumsymm{C. Vafa, Mod. Phys. Lett. A4, no. 17 (1989) 1615.}
\lref\cvlg{C. Vafa, Mod. Phys. Lett. A4, no.12 (1989) 1169.}
\lref\renormglg{B. R. Greene, C. Vafa and N. P. Warner, Nuc. Phys. B324
(1989) 371.}
\lref\lgflow{C. Vafa and N. P. Warner, Phys. Lett. B218 (1989) 51.}
\lref\hcv{S. Hamidi and C. Vafa, Nuc. Phys. B279 (1987) 465.}
\lref\vc{S. Cecotti and C. Vafa, Mod. Phys. Lett. A7 (1992) 1715.}
\lref\chiralr{W. Lerche, C. Vafa and N. P. Warner, Nuc. Phys. B324 (1989)
427.}
\lref\ewessays{E. Witten, in {\it Essays on Mirror Manifolds}, ed. by S-T
Yau, (International Press, 1992).}
\lref\phases{E. Witten, {\it Phases of N=2 Theories in Two Dimensions},
IASSNS-HEP-93/3, hep-th/9301042.}
\lref\zaz{E. Zaslow, Comm. Math. Phys. 156, (1993) 301.}
\lref\gency{P. Candelas, E. Derrick and L. Parkes, {\it Generalized
Calabi-Yau Manifolds and the Mirror of a Rigid Manifold}, hep-th/9304045.}
\lref\vbdc{V. Batyrev and L. Borisov, {\it Dual Cones and Mirror Symmetry
for Generalized Calabi-Yau Manifolds}, alg-geom/9402002; V. Batyrev, {\it Dual
Polyhedra and Mirror Symmetry for Calabi-Yau Hypersurfaces in Toric Varieties},
alg-geom/9310003.}
\lref\vbhs{V. Batyrev and D. Cox, {\it On the Hodge Structure of Projective
Hypersufaces in Toric Varieties}, alg-geom/9306011}
\lref\manin{Y. I. Manin, {\it Gauge Field Theory and Complex Geometry},
(Springer-Verlag, New York, 1980); {\it Topics in Noncommutative Geometry},
(Princeton University Press, Princeton, 1991).}
\lref\mnlect{Y. I. Manin, in {\it Lecture Notes in Mathematics, no. 1111},
(Springer-Verlag, New York, 1984). }
\lref\abg{M. F. Atiyah, R. Bott and L. G{\aa}rding, Acta. Math. 131 (1973)
145.}
\lref\griffiths{P. Griffiths and J. Harris, {\it Principles of Algebraic
Geometry}, (Wiley-Interscience, New York, 1978).}
\lref\gmm{P. S. Aspinwall and C. A. L\"{u}tken, Nuc. Phys. B353 (1991)
427.}
\lref\quantumag{P. S. Aspinwall and C. A. L\"{u}tken, Nuc. Phys. B355
(1991) 482.}
\lref\singularity{S. Cecotti, L. Girardello and A. Pasquinucci, Int. J.
Mod. Phys. A6, no. 14 (1991) 2427.}
\lref\nonperturbative{S. Cecotti, L. Girardello and A. Pasquinucci, Nuc.
Phys. B328 (1989) 701.}
\lref\landausigma{S. Cecotti, Int. J. Mod. Phys. A6, no. 10 (1990) 1749.}
\lref\geometry{S. Cecotti, Nuc. Phys. B355 (1991) 755.}
\lref\rcessays{M. Ro\u{c}ek, in {\it Essays on Mirror Manifolds}, ed. by
S-T Yau, (International Press, 1992).}
\lref\essays{{\it Essays on Mirror Manifolds}, ed. by S-T Yau,
(International Press, 1992).}
\lref\martinec{E. Martinec, Phys. Lett. B217 (1989) 431.}
\lref\massive{S.. Cecotti and C. Vafa, Mod. Phys. Lett. A7 (1992) 1715.}
\lref\dewitt{B. DeWitt, {\it Supermanifolds}, (Cambridge University Press,
Cambridge, 1984).}
\lref\nemesch{D. Nemeschansky and A. Sen, Phys. Lett. B178, no. 4 (1986)
365.}
\lref\alvarez{L. Alvarez-Gaum\'{e} and P. Ginsparg, Comm. Math. Phys. 102
(1985) 311.}
\lref\yau{S-T Yau, Proc. Natl. Acad. Sci. 74 (1977) 1798.}
\lref\dixon{L. Dixon, {\it Some World-Sheet Properties of Superstring
Compactifications, On Orbifolds and Otherwise}, Princeton Preprint
PUPT-1074 (1987).}
\lref\topft{E. Witten, Comm. Math. Phys. 117 (1988) 353; Comm. Math. Phys.
118 (1988) 411.}
\lref\ey{T. Eguchi and S-K Yang, Mod. Phys. Lett. A5 (1990) 1693.}
\lref\zum{B. Zumino, Phys. Lett. B87, (1979) 203.}
\lref\lerchesmit{W. Lerche, D-J Smit and N. P. Warner, Nuc. Phys. B372
(1992) 87. }
\lref\periodcomp{P. Berglund, E. Derrick, T. Hubsch and D.
Jan\v{c}i\'{c}, {\it On Periods for String Compactifications},
hep-th/9311143.}
\lref\holform{P. Candelas, Nuc. Phys. B298 (1988) 458.}
\lref\moregriff{P. Griffiths, Ann. Math. 90 (1969) 460, 496; J. Carlson
and P. Griffiths, in {\it Journ\'{e}es de G\'{e}ometric Alg\'{e}brique
d'Angers}, eds. A. Beauville, Sijthoff and Noordhoff, (Alphen aan den
Rijn, 1980).}
\lref\peters{C. Peters and J. Steenbrink, in {\it Progress in Mathematics
vol. 39: Classification of Algebraic and Analytic Manifolds}, ed. Ueno,
(Birk\"{a}user, Boston, 1983). }
\lref\steenbrink{J. Steenbrink, Comp. Math. 34 (1977) 211.}
\lref\dolgachev{I. Dolgachev, in {\it Lecture Notes in Mathematics 956},
eds. A. Dold and B. Echmann, (Springer-Verlag, 1982).}
\lref\pdm{P. Green and T. Hubsch, Comm. Math. Phys. 113 (1987) 505.}
\lref\verlinde{E. Verlinde and H. Verlinde, Phys. Lett. B192 (1987) 95.}
\lref\wen{X-G Wen and E. Witten, Nuc. Phys. B261 (1985) 651.}

\Title{HUTP-94/A002}{\vbox{\centerline{Supermanifolds, Rigid Manifolds and
Mirror Symmetry}}}
\centerline{S. Sethi\footnote{$^\ast$} {Supported by the Fannie and John
Hertz Foundation and the Beinecke Memorial Fellowship. }\footnote{$^\dagger
$}{(sethi@string.harvard.edu)} }

\bigskip\centerline{Lyman Laboratory of Physics}
\centerline{Harvard University}\centerline{Cambridge, MA 02138, USA}

\vskip 1in

% Abstract

By providing a general correspondence between Landau-Ginzburg orbifolds and
non-linear sigma models, we find that the elusive mirror of a rigid
manifold is actually a supermanifold. We also discuss when sigma models
with super-target spaces are conformally invariant and describe their
chiral rings. Both supermanifolds with and without \kh moduli are
considered. This work leads us to conclude that mirror symmetry should be
viewed as a relation among super-varieties rather than bosonic varieties.

\Date{4/94}
\newsec{Introduction}

Mirror symmetry is among the most exciting and intriguing discoveries in
string theory. We begin with a brief description: Consider the class of
string vacua given by ${\rm N}=2$ superconformal non-linear sigma models. These
theories are particularly interesting when used to compactify the heterotic
string; the resulting theory is then space-time supersymmetric. Mirror
symmetry is the statement that strings propagating on inequivalent but
mirror target spaces give physically equivalent theories. To understand why
a manifold and its mirror give the same field theory, we must examine how
geometric rings on the manifold map to rings of operators in the field
theory.

In particular, the cohomology ring of the manifold, in the large radius
limit, is associated to a ring of primary operators in the field theory.
However, there are two possible candidates for the cohomology ring,
excluding complex conjugation, in a unitary ${\rm N}=2$ theory: Either the ring
of left-chiral right-chiral primary operators, known as the $(c,c)$ ring,
or the ring of left-antichiral right-chiral primary operators known as the
$(a,c)$ ring \chiralr\dixon. The operators in these rings are labeled by a
left-moving and a right-moving $U(1)$ charge. Chiral primary operators are
positively charged while anti-chiral primary operators are negatively
charged. Note that the two rings are exchanged by simply changing the sign
of the left-moving $U(1)$ charge. If the cohomology ring of a manifold $
{\it M}$ is identified with the $ (c,c)$ ring, then the mirror symmetry
conjecture asserts that there is a mirror manifold $ {\it \widetilde{M}}$ whose
cohomology ring can be identified with the $ (a,c)$ ring of the field theory.
For manifolds of complex dimension $d$, the Hodge numbers $ h^{p,q}$ of $
{\it M}$ are related to the Hodge numbers $ \tilde{h}^{p,q}$ of $ {\it
\widetilde{M}}$ by:

\eqn\Hodge{\tilde{h}^{p,q}= h^{{d-p},q}  .}

There are two kinds of moduli for these conformal field theories. Those
given by charge $(1,1)$ primary operators correspond to deformations of the
\kh structure, while those given by charge $(-1,1)$ primary operators
correspond to deformations of the complex structure. The $(-1,1)$ operators
are identified with $(d-1,1)$ operators using the asymmetric spectral flow
isomorphism of these theories. Under the mirror map, these two moduli
spaces are exchanged.

Within the past few years, there has been a great deal of effort devoted to
understanding mirror symmetry \essays. A primary motivation has been to
understand how string theory modifies conventional geometry, providing
powerful and unexpected relations between a priori unrelated manifolds.
However, the existence of rigid manifolds has posed a long-standing
obstacle to this effort. A rigid manifold is a Calabi-Yau manifold without
any complex structure moduli. The mirror therefore cannot possess any \kh
moduli and so cannot be a \kh manifold in any conventional sense. In a
number of cases, the mirror to a rigid manifold can be described as a
Landau-Ginzburg orbifold \cvessays. These Landau-Ginzburg models fall
outside of the class that had previously admitted a sigma model
interpretation \renormglg\phases\martinec. Our approach to this problem is
to give a sigma model interpretation to these Landau-Ginzburg orbifolds.
These theories include a far larger class of models than simply mirrors of
rigid manifolds. We then find that strings propagating on a bosonic
manifold can be alternatively described by strings propagating on a
supermanifold where the fermionic coordinates carry negative dimension!

In fact, the fermionic coordinates perform a second crucial function by
`cancelling' out the contribution to the super-first Chern class from the
bosonic coordinates; hence, allowing conformal invariance. From this
analysis, we conclude that the correct framework for understanding mirror
symmetry is not the space of bosonic varieties but the space of
super-varieties. This framework would seem a more natural setting for
algebraic geometry since the mirror encodes non-trivial information about
the original manifold - such as the counting of instantons - in a
computable manner.

In the following section, we give a geometric interpretation of
Landau-Ginzburg orbifolds as sigma models on supermanifolds. Section three
provides a discussion of when such sigma models are conformally invariant.
Section four is a study of the chiral primary ring of these theories.
This section includes a discussion of the type of supercohomology theory
needed
to construct physical observables. The final section is devoted to
summarizing the findings.

\newsec{Landau-Ginzburg Orbifolds and Non-linear Sigma Models}
\subsec{A Path-Integral Argument} The action for a two-dimensional $ {\rm N}=2$
\lg model is of the form:

\eqn\action{\int{d^2 z d^4 \theta K(\Phi_i, \Phi_{\bar{i }})} + ( \int{d^2
z d^2 \theta^- W(\Phi_i)} + c.c.) . } The chiral superfields $ \Phi^i$
satisfy $ D^+ \Phi^i = \overline{D}^+ \Phi^i = 0$ where

\eqn\define{D^\pm = {\partial \over \partial \theta^\pm}{ }+ \theta^\mp
{\partial \over \partial z}{ } \qquad \qquad\theta^\pm \buildrel
c.c.\over\longrightarrow \overline{\theta}^\mp ,}
while the anti-chiral fields satisfy conjugate conditions. We take the
superpotential
to be a quasi-homogeneous polynomial in the chiral superfields with a
degenerate critical
point at the origin. A simple example is the Fermat type superpotential
where the fields $ \Phi_i$ have charge $( {1 \over k_i}, {1 \over k_i})$ and

\eqn\superpot{W(\Phi_i) = \sum{{\Phi_i}^{k_i}} .}

Let us denote the charge of the chiral field $ \Phi_i$ by $( {1 \over k_i},
{1 \over k_i})$ for the general case. The requirement of quasi-homogeneity
is needed for
conformal invariance. The superpotential then defines a universality
class under renormalization group flow. The choice of superpotential
determines the chiral
primary ring $ \ch$ in a simple way \chiralr\lgflow ,

\eqn\chiralring{\ch = {{\bf C}[\Phi_i] \over dW(\Phi_i)}. }
This is the local ring of $W$ given by monomials in $\Phi_{i}$ modulo
the Jacobian ideal. To simplify our discussion, we take
the low-energy limit and consider constant superfields which we can treat
as complex variables. A computation of $ \c = {c \over 3}$, where $ c$ is
the central charge, gives the well-known formula $ \c = \sum{1-{2 \over
{k_i}}}$. This is the dimension of the associated sigma model. Our aim in
this section is to relate orbifolds of \lg models with integral $ \c$ to
sigma models. Specifically, the orbifold of the \lg model by the diagonal
sub-group of the phase symmetries for the theory.

\eqn\phase{\Phi_i\longrightarrow e^{2 \pi i \over k_i} \Phi_i }

Let us briefly comment on what we mean by associating a sigma model to a
\lg orbifold. The \lg theory and the sigma model can be viewed as different
`phases' of the same theory. A smooth analytic continuation is
believed to exist between the two phases \phases\vc. For the remainder of
this
paper, we simply refer to an identification or association of a sigma model
to a \lg orbifold with the above comments in mind.

Before discussing the general case, let us consider a simple but
illuminating family of examples with superpotential:

\eqn\ex{W(\Phi_i) = \sum_{i=1}^{3 N}{{\Phi_i}^3} .} For the case $ N=1$, we
can use the results of \renormglg. Let $ \Lambda = {\Phi_1}^3$ and $ z_i =
{\Phi_i \over \Phi_1}$ for $ i \ne 1$. This corresponds to choosing a patch
with $ \Phi_1 \ne 0$. The F-term in the lagrangian {\action} then becomes:

\eqn\Fterm{\int d^2z d^2\theta^- \Lambda (1+z_2^3+z_3^3) + c.c. }
 More importantly, the
Jacobian for this change of variables is trivial so we can integrate out
the superfield $\Lambda$ as a Lagrange multiplier giving a super-delta
function. The change of variables is not one-to-one, so we must further
orbifold by the diagonal $ {\bf Z}_3$ sub-group of the phase symmetries. We
conclude that the \lg orbifold is identified with a sigma model on the
variety $ W=0$ in $ {\bf P}^2$, which is just the maximal $ SU(3)$ torus.

For the more general superpotential with $ n$ fields, this change of
variables procedure can be performed when $ \c = n-2$. This condition
corresponds to the variety $ W=0$ in weighted projective space having
vanishing first chern class. There are two other cases that can arise. For
$ \c > n-2$, one simply adds quadratic fields to the superpotential $ W
\rightarrow W+x_1^2+\ldots$ until $ \c = n-2$. The quadratic fields have no
effect on the chiral ring or the conformal fixed point to which the \lg
theory flows.

The more interesting case is $ \c < n-2$ which includes \ex\ for $ N>1$. We
proceed with the following ansatz: Add bilinears of ghost superfields to
the superpotential $ W \rightarrow W+\gh_1 \gh_2+\ldots$. By ghost
superfields $ \gh_i$, we mean each component of the superfield has
reversed statistics so the lowest component is a spin zero fermion etc.
Under a change of variables, the ghost measure transforms inversely to the
measure for the bosonic superfields. By adding enough pairs of ghosts, the
Jacobian under the change of variables procedure can be made trivial. In
addition, there is again no change in the chiral ring for the theory so we
expect the conformal \lg theory to be unchanged. Under the condition that $
\c $ be integral, the sum of the charges $ \sum{1 \over k_i}$ for theories
in this class can be integral or half-integral. The first case only
requires adding ghost bilinears while the second needs a quadratic bosonic
field as well as ghost bilinears. The requirement that the Jacobian be
trivial and that the sigma model have dimension $ \c$ uniquely determines,
for $ W$ with degree $ > 2$ (the nontrivial case), the number of ghost and
quadratic bosonic fields needed for this procedure.

Let us return to our examples \ex\ for $ N>1$. For these cases, the
dimension computed from the \lg model gives $ \c = N$. We claim that
orbifolding by the diagonal phase group identifies these models with sigma
models on target spaces defined by the zero set of the algebraic
constraints:

\eqna\super $$\eqalignno{ N &{\rm = 2 \qquad\qquad
\sum_{i=1}^{6}{z_i^3+\gh_1\gh_2}} &\super a\cr N &{\rm = 3 \qquad\qquad
\sum_{i=1}^{9}{z_i^3+\gh_1\gh_2+\gh_3\gh_4}} &\super b\cr \vdots
&{\qquad\qquad\qquad\quad\vdots}\cr}$$ These constraints define
supermanifolds - the subject of the following section - in super-projective
space. Note that in the sigma model, the ghost superfields carry negative
quantum dimension since they contribute to $ \c$ with opposite sign to
bosonic superfields. In the \lg theory, however, they have no effect on $
\c$. That the dimensions of both models agree is an encouraging first sign.

\subsec{Relation to Rigid Manifolds}

Using standard techniques \cvlg, we can compute the Hodge diamonds for the
orbifolds of the \lg models \ex.\ These are displayed for the first two
cases in figure $ 2.1$.

$$\matrix{ & & & & & \qquad & & & & 1 & & & \cr & & 1 & & &\qquad & & & 0 &
& 0 & & \cr & 0 & & 0 & &\qquad & & 0 & & 0 & & 0 & \cr 1 & & 20 & & 1
&\qquad & 1 & & 84 & & 84 & & 1\cr & 0 & & 0 & &\qquad & & 0 & & 0 & & 0 &
\cr & & 1 & & &\qquad & & & 0 & & 0 & & \cr & & & & & \qquad & & & & 1 & &
& \cr }$$

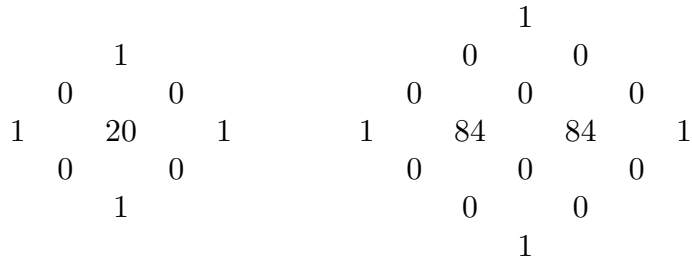
\noindent {\ninerm \it Figure 2.1: Hodge diamonds for the $ N=2$ and
$ N=3$ cases.}
\vskip 0.1in

For the case $ N>2$, we find that the models have no \kh
deformations. We therefore expect these models to be mirrors of rigid
manifolds. Indeed the mirrors for these \lg orbifolds with $ N>2$ are rigid
toroidal orbifolds \gmm. The mirror for the $ N=2$ case is the $ K3$
surface. The supermanifolds defined by \super{a, \ldots}\ should therefore
be the elusive geometric mirrors. Many \lg orbifolds falling into this
class of theories do, however, possess \kh forms which are contributed
from the
twisted sectors.

The more general superpotential can contain quartic and higher terms in the
ghost superfields. In these cases, the fermionic geometry in the \lg model
is nontrivial. There can also be mixed terms containing both bosonic and
fermionic fields such as $ z_1 \gh_1\gh_2$. However, since
the ghost fields now appear nontrivially in the chiral ring for these \lg
theories, we believe these models are generally nonunitary.

\newsec{Supermanifolds}
\subsec{Conditions for Conformal Invariance}

Homogeneous coordinates for superprojective space $ \sp(n|m)$ are given by

\eqn\coordinates{(z^1,\ldots,z^{n+1}|\gh^1,\ldots,\gh^m)}
with $ z^\alpha\sim\lambda
z^\alpha$ and $ \gh^\alpha\sim\lambda \gh^\alpha$. There are $ n+1$
coordinate patches with $ z^i \ne 0$ in the $ i$-th patch.  Inhomogeneous
coordinates are then
defined in the standard manner $ (\tilde{z}^1,\ldots,\hat{\tilde{z}^i},
\ldots, \tilde{z}^{n+1}|\tilde{\gh}^1,\ldots,\tilde{\gh}^m)$ where $
\tilde{z}^j ={ z^j\over z^i}$ and $\tilde{\gh}^j = {\gh^j \over z^i} $. Let
$ \tilde{z}^j_{\{k\}}$ and $ \tilde{\gh}^j_{\{k\}} $ be coordinates in the
patch where $ z_k \ne 0$, then the transition functions are the usual ones:
$ \tilde{z}^j_{\{k\}} = \tilde{z}^j_{\{i\}} {z^i \over z^k}$ and $
\tilde{\gh}^j_{\{k\}} = \tilde{\gh}^j_{\{i\}} {z^i \over z^k}$. The
fermionic coordinates $\tilde{\gh}^j $ are Grassmann-valued sections of the
line bundle $ O_{\bf P^n}(-1)$. This space is a split
supermanifold\foot{Split supermanifolds are also known as DeWitt
supermanifolds \dewitt. The de Rham cohomology for the split case reduces
to
the usual de Rham cohomology of the body, in this case $ \bf P^n$, so these
spaces are quite uninteresting cohomologically.} - a special case of the
super-Grassmannian \manin. The weighted case $
\wsp(k_1,\ldots,k_{n+1}|l_1,\ldots,l_m) $ is a straightforward extension
with $ z^\alpha\sim\lambda^{k_\alpha} z^\alpha$ and $
\gh^\alpha\sim\lambda^{l_\alpha} \gh^\alpha$.

We are interested in $ {\rm N}=2$ sigma models on subvarieties of these spaces
with actions of the form:

\eqn\sigmaaction{\int{d^2 z d^4 \theta K(\Phi^\alpha, \Phi^{\bar{\alpha}})}
+ ( \int{d^2 z d^2 \theta^- \Lambda P(\Phi^\alpha)} + c.c.). } Let the
indices $ \mu,\nu,\ldots$ refer to bosonic coordinates, $ i,j,\ldots$ to
ghost coordinates, and $ \alpha,\beta,\ldots$ for either case. Let us
establish a convention for the component expansion of the superfields:
\eqn\comp{\eqalign{\Phi^\mu &= z^\mu - \theta^- \psi_{+}^\mu -
\bar{\theta}^- \psi_{-}^\mu + \theta^-\bar{\theta}^- F^\mu - \ldots \cr
\Phi^i &= \gh^i - \theta^- \xi_{+}^i - \bar{\theta}^- \xi_{-}^i +
\theta^-\bar{\theta}^- G^i - \ldots \cr }}

More general actions can include twisted-chiral fields \rcessays, gauge
symmetries, several polynomial constraints and both fermionic and bosonic
Lagrange multipliers. For simplicity, we restrict this discussion to the
single polynomial case with a bosonic Lagrange multiplier $ \Lambda$ and no
twisted-chiral fields. This case includes most of the interesting examples
and the extension to the other cases is not difficult.

Rather than consider the action \sigmaaction\ with an explicit F-term, let
us assume the \kh potential for the subvariety is known. We can then study
the ultraviolet divergence structure for the sigma model defined only by a
D-term. All computations can be performed in the $ {\rm N}=2$ superspace
framework, but not while preserving manifest covariance \alvarez.
Nevertheless, the final expressions for the counterterms, which are all
corrections to the \kh potential, are covariant. In the usual bosonic case,
the divergent part of the one-loop effective action is proportional to the
determinant of the \kh metric. We choose to expand the \kh potential in the
following way

\eqn\exp{K = K(\Phi^\alpha_o, \Phi^{\bar{\beta}}_o) +\tilde{ \Phi}^\alpha
\met \tilde{\Phi}^{\bar{\beta}} + \ldots } around a classical background $
\Phi^\alpha_o$ where $ \Phi^\alpha =\Phi^\alpha_o + \tilde{ \Phi}^\alpha$.
The order of the terms is important since the fields can anticommute. The
supermetric is defined to be $\met$ . The divergent part of the one-loop
counter-term is now proportional to $ \ln \sdet \met$. The condition for
conformal invariance to one-loop is the existence of a super-Ricci flat
metric. Standard arguments imply that all higher loop counterterms are
cohomologically trivial. In fact, the actual metric for the conformal
theory theory will not be the super-Ricci flat metric but a metric
corrected for higher loop contributions as in the usual bosonic case
\nemesch.

\subsec{Super-Ricci Flat Metrics}

Let us consider the case of $ \sp(n|n+1)$ with \kh potential

\eqn\spex{\eqalign{ K&=\ln(1+z^\alpha z_\alpha) \qquad z_\alpha =
\delta_{\alpha{\bar \alpha}} z^{\bar \alpha}\cr K&=\ln(1+z^\mu z_\mu) +
\sum_{p}{{(-1)^{p+1} (\gh^{i} \gh_{i})^p \over p (1+z^\mu z_\mu)^p}}\cr } }
which is a natural extension of the Fubini-Study \kh potential to
superprojective space. Surprisingly, the metric for this \kh potential is
super-Ricci flat. Unlike the purely bosonic case, this embedding space
provides a nonunitary $ \c=-1$ conformal field theory! Note that the
contribution from the ghosts to \spex\ is in the form of a globally defined
section-valued function on $ \bf P^n$. In this sense, the addition of the
ghosts to the potential does not effect the cohomology class of the \kh
form. For the more general case $ \sp(n|m)$, the \kh potential $ K =
\ln(1+z^\alpha z_\alpha)$ gives a K\"{a}hler-Einstein metric. The one-loop
counterterm, which gives the potential for the super-Ricci tensor, is in
this case:

\eqn\kepot{ \ln \sdet \met = - (n-m+1) K.} From this example, we can see
that the fermionic coordinates contribute to the super-first Chern class
with a negative sign.

Yau's theorem \yau\ for bosonic manifolds guarantees that the vanishing of
the first Chern class is necessary and sufficient to ensure the existence
of a Ricci flat metric. In the super case, there is no such theorem yet,
and in fact the situation is more subtle. We will not attempt a full
analysis of the necessary conditions for the existence of a super-Ricci
flat metric here, but we will present a preliminary investigation.

Let us examine the structure of the counterterm $ \ln \sdet \met$ more
closely. Let $ \invmet$ denote the inverse to the non-degenerate metric $
\met$ then
\eqn\structure{\ln \sdet \met = \ln \det (K_{\mu{\bar \nu}} - K_{\mu{\bar
i}}K^{{\bar i}j}K_{j{\bar \nu}} ) - \ln \det(K_{i{\bar j}}).} For metric
non-degeneracy, terms quadratic in the ghost fields must be present in the
\kh potential. These terms ensure that $ K_{i{\bar j}}$ is invertible, and
are also necessary if the bosonic part of this counterterm is to be
cohomologically trivial. Physically, these terms ensure that the ghost
field propagators are well-defined. The contribution to the fermionic part
$ K_{ferm}$ of the \kh potential for weighted projective space $
\wsp(1,\ldots,1|l_1,\ldots,l_m) $ should take the form:
\eqn\kferm{K_{ferm} = \sum{\gh^i \gh_i \over (1+z^\mu z_\mu)^{l_i}} +
O(\gh^i \gh_i)^2 .} The bosonic part of the \kh potential is taken to be the
usual Fubini-Study potential. This particular embedding space is important
for understanding the examples given in \super{a,\ldots}\ and we will infer
general features from this case. Note that if the bosonic coordinates scale
with nontrivial weight, then a non-degenerate (weighted) extension of this
Fubini-Study potential does not generally exist. To ensure the bosonic part
of the counterterm \structure\ is trivial, the super-first Chern class must
vanish. Fermions of weight $ l_i$ contribute $ -l_i$ to the super-first
Chern class. Unlike the bosonic case, however, vanishing of the super-first
Chern class is not sufficient to ensure the existence of a nontrivial
super-Ricci flat metric. Physically, $ K_{ferm}$ can still be renormalized
from the one-loop and higher counterterms since the terms quadratic in the
ghosts are globally defined. For example, the target space $ \sp(1,1|2)$
with bosonic field $ z$, ghost field $ \gh$ and potential
\eqn\trex{K = \ln(1+z {\bar z}) +{ \gh {\bar \gh} \over (1+z {\bar z})^2}}
has a counterterm $ \propto {2 \gh {\bar \gh} \over (1+z {\bar z})^2}$. The
theory therefore flows to a conformal model with a degenerate metric. If a
non-degenerate metric exists for a variety in $
\wsp(k_1,\ldots,k_{n+1}|l_1,\ldots,l_m) $ with $\sum{k_i} - \sum{l_j} = 0
$, and there is no renormalization of  \kh potential terms quadratic in the
ghost fields, then the theory flows to a nontrivial super-Ricci flat
metric.

For those sigma models admitting a \lg description, we fully expect from
the arguments given in section two that a nontrivial super-Ricci flat
metric exists. As a basic check, we can compute the super-first Chern class
for \lg orbifolds corresponding to supermanifolds. Recall that each
superfield $ \Phi_i$ has charge $( {1 \over k_i}, {1 \over k_i})$. Each of
the $ (\sum{{1 \over k_i} -1)}$ ghost bilinears then subtracts $ d$ from
the super-first Chern class. The degree $ d$ of $ W$ is the homogeneity of
$ W$ as a defining equation in weighted projective space. The bosonic
fields contribute $ d \over k_i$ each while the constraint $ W$ further
reduces the super-first Chern class by $ d$. Just as we expect, the
super-first Chern class vanishes. The change of variables procedure is
therefore only possible when the super-variety $ W=0$ has vanishing
super-first Chern class. A more direct computation from the action
\sigmaaction\ using the constraint term shows that the induced super-Ricci
tensors for the models \super\, are indeed cohomologically trivial.

Within this framework, we can now interpret the models \super\, as
varieties in $ \wsp(\underbrace{1,\ldots,1}_{3N}|\underbrace{1,2,\ldots,
1,2}_{N-1\, pairs})$; the $ N=2$ case is a $ (4|2)$ supermanifold while the
$ N=3$ case is a $ (7|4)$ supermanifold. For these models, each ghost bilinear
in the defining polynomial scales as $ \lambda^3$. The assigment of scaling
weight to each ghost field is therefore unique in these cases. For
polynomials of higher degree $ d$ derived from \lg theories, there is some
freedom in the assignment of scaling weight to each ghost field. However,
not all choices admit non-degenerate super-Ricci flat metrics. Further, the
chiral rings can differ for different choices of scaling weights. These
considerations limit the freedom in choosing ghost scaling weights when
identifying \lg orbifolds with supermanifolds.

\newsec{The Chiral Ring}

This section is organized in the following way: We first examine general
features of these theories. This discussion is in the context of the
topological sigma model. We then proceed to study the structure of differential
forms on the body of the supermanifold. Guided by that
analysis, we conjecture the form of an analogue to the usual holomorphic
$n$-form and study the `variation of Hodge structure.' In this way, we
construct the local observables for the sigma model phase. Lastly, we
consider the cases with \kh moduli and explain how these deformations can
arise.

\subsec{General Features}

There are two conserved $U(1)$ charges for an ${\rm N}=2$ superconformal model.
Let us list the left and right charge decomposition of the operators
pertinent to this discussion. The fields $(\psi^\alpha_{+},
\psi^{\bar\alpha}_{-})$ are assigned charge $(1,0)$ and $ (0,1)$, $
(\psi^\alpha_{-}, \psi^{\bar\alpha}_{+})$ charge $ (-1,0)$ and $
(0,-1)$ respectively. The fields $(z^\alpha, z^{\bar\alpha})$ are
uncharged. There are also two supercharges $G^\pm$ with charge $(\pm
1,0)$ together with their conjugates $\bar{G}^\mp$. The topological
sigma model
(A-model) is obtained by twisting the $ {\rm N}=2$ theory. This is accomplished
by coupling the vector $ U(1)$ current to a background gauge field $ A$
\topft\ey,

\eqn\twisting{ S\rightarrow S + \int{{\bar J}A + J {\bar A}}.} The
background gauge field is taken to be one-half the spin connection. The
spins of the fields in the topological theory are determined by the shifted
stress-energy tensor:

\eqn\stress{T\rightarrow T - {1\over 2} \partial J \qquad {\bar
T}\rightarrow{\bar T}+{1\over 2} {\bar\partial}{\bar J}.}
In particular, two of the supersymmetry generators, $G^+$
and $ {\bar G}^-$, are now spin zero. These nilpotent charges are
interpreted
as generators of BRST transformations. For this choice of twisting,
elements of the $ (c,c)$ ring are identified with BRST cohomology classes
of the operator $Q = G^+ + {\bar G}^- $.

Let us use $\sm$ to denote the target supermanifold. The body of $\sm$ is
a \kh
manifold with $c_1 > 0$ and dimension $\b > \c$. On a genus $g$
Riemann surface $\Sigma$, the twisted theory has a background charge
given by $\c \, (1-g, 1-g)$. Since we are studying conformal models, this
background charge violation is independent of the map $z:
\Sigma \rightarrow \sm$. This is another way of saying that the axial
$U(1)$ charge is conserved. Therefore correlation functions $ <\vartheta>$
vanish
unless the operator $ \vartheta$ has charge $ (\c, \c)$ for $\Sigma$
genus $0$. The genus zero correlation functions are of particular
interest since they are related to Yukawa couplings for the low energy
theory.

In the usual case where $\sm$ is a bosonic target space, the local
observables for the A-model are identified with elements of the de Rham
cohomology for $\sm$ \ewessays. The observables are constructed from the
fields $(\psi^\mu_{+},\psi^{\bar\mu}_{-})$ which we call
$(\psi^\mu,\psi^{\bar\mu})$
for simplicity. To a differential form $\omega$ on $\sm$,

\eqn\diffform{\omega = \omega_{{\mu_1}\ldots {\mu_p}{{\bar\mu}_1}\ldots
{{\bar\mu}_q}} dz^{\mu_1} \wedge\cdots\wedge dz^{{\bar\mu}_q},}
we associate an operator $\vartheta_\omega$ in the field theory by making
the substitution:

\eqn\substitute{dz^{\mu}\rightarrow \psi^\mu \qquad\qquad
dz^{\bar\mu}\rightarrow
\psi^{\bar\mu}.}

The action of $Q$ on $\vartheta_\omega$ is just that of the exterior
derivative $d$ on $\omega$. When $\sm$ is a supermanifold, supersymmetry
again provides us with an exterior derivative.
\eqn\exterior{\eqalign{& d= \partial + {\bar \partial} \cr &d = \psi^\mu
{\partial
\over \partial z^\mu} + \xi^i {\partial \over \partial \eta^i} + c.c.\cr}}
The observables are again in the cohomology of this nilpotent operator.
However, there are immediate problems if we na\"{\i}vely try to identify
observables with `forms' on $\sm$ by substituting the bosonic field $\xi$
for $d\eta$ as above.
Let us momentarily  restrict to the constant maps $z:
\Sigma \rightarrow \sm$. Integration over the moduli space of these maps
is simply integration over the target space. For an $(n|m)$
supermanifold, there are $n$ zero modes for $\psi$ together with $m$ zero
modes for $\xi$. Firstly, when evaluating correlation functions,
integration over the
tangent vectors $\xi$ to the ghost directions yields infinities.
Secondly, the fields $\xi$ are positively charged but adding the ghost
coordinates {\it decreases } $\c$. From these considerations, we
should expect to identify $d\eta$ with a negatively charged operator.
Otherwise, all
correlators vanish by charge conservation! Lastly, viewing $d\eta$ as a
measure
for Berezin integration requires the operator identified with $d\eta$ to
transform
inversely to $\eta$ under a change of coordinates -- unlike $\xi$. We will
argue later
that distribution-valued forms resolve these issues with the
identification:

\eqn\identify{d\eta^i \rightarrow \delta (\xi^i)\qquad\qquad d\eta^{\bar
i} \rightarrow
\delta (\xi^{\bar i}).}

At first glance, it appears that the sigma models we are studying must
possess a \kh modulus. After all, there is a natural \kh form induced from
the embedding space:

\eqn\kahler{k = \partial {\bar \partial} K.}
Indeed, it was shown long ago that ${\rm N}=2$ supersymmetry requires a \kh
target
space \zum. Does this contradict our asserted correspondence between \lg
orbifolds and these sigma models? How can the mirror of a rigid manifold
have a \kh modulus? Let us consider correlation functions
involving the \kh form. The natural volume form for absorbing the
fermionic zero modes $(\psi^{\mu},\psi^{\bar\mu})$ is $k^{\b}$, but
$<k^{\b}>$ vanishes
by charge conservation. Taking higher powers of $k$ only worsens the
situation. Clearly, the natural volume form for the supermanifold is not
constructed from the induced \kh form. That the ${\overline{27}}^3$ Yukawa
coupling
for the low-energy string theory vanishes is exactly the behavior
expected from mirror symmetry. There should be no marginal perturbation
associated with the \kh parameter.  However, this is not
sufficient. We must further show that the correlator $<k^n \vartheta>$
vanishes for any observable $\vartheta$ with $n>0$. This condition is the
requirement that the \kh form completely decouple from the chiral ring.
We will return to check this condition after studying the properties of
`good'
observables for $\sm$.

Usually, the ring of observables for the A-model is sensitive to rational
curves
on the target space. Let us show that when the target is a supermanifold,
there is no such
sensitivity. Path-integral computations in the A-model therefore reduce to
integration over the space of
constant maps. For simplicity, let us take maps $z: \sp (1|0)
\rightarrow \sm$ where $\sm$ is a variety in $\sp (n|m)$ defined by a
degree $q$ constraint. Note that $q < n+1$ for conformal invariance. As a
warmup, let us first consider constant maps. Maps into $\sp (n|m)$ can be
parametrized by $(a^0,\ldots,a^n, \tilde{a}^1,\ldots,\tilde{a}^m)$ with a
single scaling relation: $a^j\sim\lambda a^j$ and $\tilde{a}^i\sim\lambda
\tilde{a}^i$. The $a^j$ are bosonic while the $\tilde{a}^i$ are
fermionic. This is just a copy of $\sp (n|m)$ and enforcing the
constraint reduces the moduli space to a copy of $\sm$. Let $(x,y)$
denote homogeneous coordinates for $\sp (1|0)$. Take the coordinates
$z$ for $\sp (n|m)$ to be sections of $O(k)$ over $\sp (1|0)$.

\eqn\sections{\matrix{&  & \qquad & \cr
                      & z^j = \sum a^{j}_{l} x^l y^{k-l} & \qquad & 0 \leq j
\leq n\cr
                      &  & \qquad & \cr
                      & \eta^i = \sum \tilde{a}^{i}_{l} x^l y^{k-l}& \qquad &
1\leq i \leq
                      m \cr
                      &  & \qquad & \cr}}
This space is a copy of $\sp ( \{ n+1 \} \{k+1\} - 1 | m\{k+1\} )$ where
we have implicitly compactified the moduli space. The constraint reduces
the bosonic dimension of the moduli space by $kq+1$ to $k(n+1-q)+(n-1)$
in the generic case. This is to be contrasted with the Calabi-Yau case
where the dimension of the moduli space is independent of  $ k$. The
additional fermionic
tangent vectors to the moduli space then annihilate correlation functions
for $k>0$.  This argument extends straightforwardly to varieties in
weighted projective space. The decoupling phenomena described above occurs
generally in these models:
Topological amplitudes are therefore independent of perturbations of the
induced \kh form.

\subsec{The Hodge Structure}

In the previous section, we argued that there is no
\kh modulus when $W=0$ is a smooth variety. The interesting
moduli are then derived from the $(a,c)$ ring, and are related to
deformations of the complex structure. Let us establish some notation.
In the \lg theory, the defining superpotential $W(z^1,\ldots,z^{n+1})$
of degree $d$ can be taken to be ghost free for unitary models. Let $p =
W + \Gamma$ denote the superconstraint where

\eqn\superconstraint{\Gamma = \eta^1\eta^2 + \ldots + \eta^{2m-1}\eta^{2m},}
and $\Gamma^{m+1} = 0$. Let $\lambda_i =
{d\over k_i}$ be the weight of  $z^i$ , and ${\tilde\lambda}_i$ be the
weight of $\eta^i$. The supermanifold $\sm$ is then a degree $d$ variety in
$\wsp (\lambda_1,\ldots ,\lambda_{n+1}|{\tilde\lambda}_1,\ldots
,{\tilde\lambda}_{2m})$. Let us briefly review the structure of the
chiral ring expected from spectral flow arguments. This discussion applies
generally
to ${\rm N}=2$ superconformal theories with integral $U(1)$ charges in the
Neveu-Schwarz sector. The left-moving $U(1)$ current can be expressed in
bosonized form as\foot{See \chiralr\ for an explanation of the
normalization conventions, and a more detailed discussion.}

\eqn\bosonized{J = i \sqrt{\c} \, \partial \phi_L ,}
and similarly for the right-mover. The unique states with charge $(\c,0)$
and $(0,\c)$ are constructed from the free bosons $\phi_L$ and $\phi_R$:

\eqn\forms{\Omega = e^{i\sqrt{\c}\, \phi_L} \qquad\qquad
           {\bar\Omega} = e^{-i\sqrt{\c}\, \phi_R}.}
As the notation suggests, these operators correspond to the
holomorphic and anti-holomorphic forms present on a Calabi-Yau manifold
$\cy$. The operator corresponding to the volume form is just
$\Omega\otimes{\bar\Omega}$. States in the $(a,c)$ ring with charge
$(-p,q)$ flow to states $(\c-p,q)$ in the $(c,c)$ ring under spectral
flow generated by $\Omega$. The corresponding geometric operation on the
Calabi-Yau is the cup product of an element in $H^q (\cy, \wedge^p {\it
T}\cy)$ with the holomorphic $\c$-form. For theories with integral $U(1)$
charges, spectral flow provides a precise correspondence between the
$(a,c)$ and $(c,c)$ rings.

By computing the $(c,c)$ ring for the \lg orbifold, we know the expected
structure of the $(c,c)$ ring for $\sm$. Unfortunately, there is no
current supercohomology theory that would provide the desired ring, or
even the correct Hodge numbers. We must therefore proceed to construct
the observables for the sigma model guided by physical considerations.

Let us return momentarily to the family of examples \super{}. It is
instructive to examine the Hodge diamonds of the bodies $\cy$ of these
supermanifolds. A straightforward computation reveals that
the Hodge numbers $h^{p,q}$ of middle cohomology ($p+q=\c$) of the \lg orbifold
agree with $h^{p+m,q+m}_{o}$ of $\cy$. The subscripted Hodge number
$h^{p,q}_{o}$ refers to the number of primitive forms in the Dolbeault
group $H^{p,q}(\cy)$ \griffiths. As an illustration, compare figure $4.1$
with figure $2.1$ showing the $N=2$ $(m=1)$ case.
\vskip 0.2in
$$\matrix{ & & & & &1 & & & & \cr
           & & & &0 & &0 & & & \cr
           & & &0 & &1 & &0 & & \cr
           & &0 & &0 & &0 & &0 & \cr
           &0 & &1 & &21 & &1 & &0 \cr
           & &0 & &0 & &0 & &0 &\cr
           & & &0 & &1 & &0 & & \cr
           & & & &0 & &0 & & & \cr
            & & & & &1 & & & & \cr}$$

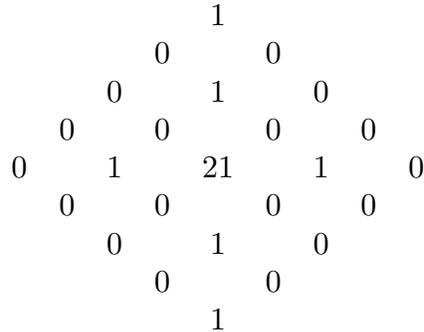
\noindent {\ninerm \it Figure 4.1: Hodge diamond for the body of the $N=2$
case.}
\vskip 0.2in
The agreement of these Hodge numbers is a consequence of the
\lg description of these models. To explain this point, let us describe
the construction of middle cohomology on the body $\cy$ defined by $W=0$.
The main theme of this discussion is the relation between pole-order and
charge grading of differential forms. Let $\emb$ denote the embedding
space ${\bf WP}(\lambda_1,\ldots ,\lambda_{n+1})$. The Poincar\'{e} residue
of a holomorphic form $\varpi$ on $\emb$ with a pole on $\cy$ is given by
\moregriff\peters\ :

\eqn\residue{\res [\varpi ] = \int_{\gamma} \varpi .}
The integration contour is a small one-cycle enclosing the hypersurface
$\cy$. For a cycle $\gamma \in H_{k-1}(\cy)$, the residue satisfies

\eqn\tubes{\int_{\gamma} \res [\varpi] = {1\over 2\pi i} \int_{T(\gamma)}
\varpi}
where  $T(\gamma) \in H_{k}(\emb - \cy)$ is a tube over $\gamma$. As
an example, let us take $\cy$ to be Calabi-Yau. The holomorphic $n$-form
$\Omega$
then has the following well-known construction \abg\holform:

\eqn\calabiyau{\Omega = \res [{\omega\over W}] \qquad \omega = \sum_{i}^{n+1}
 (-1)^{i} \lambda_{i}\, z^{i} dz^{1}\ldots\wedge \hat{dz^i} \ldots \wedge
 dz^{n+1}.}
This construction is only well-defined in the Calabi-Yau case where the
scaling degree of $W$ equals that of $\omega$. More
generally, the pole-order of the form on $\emb$ determines the Hodge
decomposition
of the form on $\cy$ obtained under the residue map. To construct the
middle cohomology of $\cy$, take the form

\eqn\middle{\varpi (P) = {P(z^1,\ldots,z^{n+1})\, \omega \over W^k},}
with $\omega$ defined in \calabiyau. This is only well-defined when the
condition

\eqn\welldefined{k\, \deg (W) = \deg (P) + \sum \lambda_{i}}
is satisfied. Under the Poincar\'{e} residue, this form maps into
 $\oplus_{q=1}^{k} H_{o}^{n-q, q-1} (\cy)$. Returning to the \lg theory, recall
the definition of the chiral primary ring $\ch$ given in \chiralring. After
orbifolding, only the ring elements with integral charge survive, so we
restrict
our discussion to that subring. Let  $P_\alpha$ be a basis for $\ch$, then
$\res [\varpi (P_\alpha)]$ is a basis for $H_{o}^{n-1} (\cy)$
\moregriff\steenbrink\dolgachev. More precisely, if $P\in\ch$ satisfies
\welldefined\ for some $k$, then $\res [\varpi (P)] \in H_{o}^{n-k,k-1}(\cy)$
is nontrivial. This description of the Hodge structure has recently been
extended
to hypersurfaces in toric varieties \vbhs. For the concrete case of \ex\
with $N=2$, the forms

\eqn\formex{ \left( \,{\omega\over W^2}, \, {(z^i z^j z^k)\, \omega\over W^3}
\,
(i\not= j \not= k),\, {(\prod z^i) \,\omega\over W^4}\,\right) }
provide a basis under the residue map for the middle cohomology shown in
figure $4.1$. In this way, we
obtain a map from $\ch$ to the primitive cohomology of the body of $\sm$.

This analysis is quite independent of any sigma
model interpretation for the \lg orbifold. It was shown in \landausigma\
that this mapping of Hodge structures is consistent with the real
structure and period maps. Computing the chiral ring structure constants
on the body of $\sm$ produces the same results as computations for the
\lg orbifold including normalization \landausigma\geometry. Picard-Fuchs
equations can therefore be derived for general \lg orbifolds with integral
$\c$. Direct computations of period matrices can also be performed
for these models \lerchesmit\periodcomp. Using these techniques, the
dependence of the \lg orbifold on the complex structure moduli can be
recovered.\foot{At least for complex structure moduli that can be parametrized
by
polynomial deformations  \pdm. This is not true for more general cases such
as embeddings in products of super-projective spaces. Even an embedding in a
single
weighted projective space can produce moduli not parametrized by
polynomial deformations. From the \lg viewpoint, there are additional moduli
contributed from the twisted sectors. See \periodcomp\ for
an attempt to deal with these extra deformations.} For an
orbifold of \ex\ with $N=3$, explicit computations using these techniques
were checked against known results on the mirror manifold in \gency. In the
framework of mirror symmetry, the construction proposed in \vbdc\ provides a
general technique to obtain the body of the desired supermanifold when the
mirror is a toric variety.

The identity operator in the ring $\ch$ corresponds to the holomorphic
$\c$-form for the \lg orbifold. Under the mapping described above, the
identity maps to a $(\c+m,m)$ form. This correspondence is only natural
in the sense of preserving charge when $\cy$ is Calabi-Yau. Note that the
primitive forms described above are {\it not} observables of the sigma model.
The expected map from $\ch$ to the sigma model should be a morphism of Hodge
structures of type $(0,0)$ i.e. there should be no shift of the charges.
However,
after integrating out the ghost fields, we expect correlation functions to
reduce
to intersection theory on the body.

Let us construct the cohomology of $\sm$
excluding contributions from any fixed point sets. Those contributions are
discussed in the following subsection. The spectral flow arguments are
unaffected by the addition of the ghost coordinates. However, the
expression for $\Omega$ given in \forms\ is helpful. Take flat superspace
with \kh potential

\eqn\flatspace{K = z^\mu z_\mu + \eta^i \eta_i}
as an example. The $U(1)$ current is then a sum of $b-c$ and
$\beta-\gamma$ systems. Bosonizing each $b-c$ system provides the
identification

\eqn\bosonbc{e^{i\phi_L^\mu} = \psi^\mu .}
However, bosonizing each $\beta-\gamma$ system gives the relation
\verlinde

\eqn\bosonbg{e^{-\phi_L^i} = \delta (\xi^i).}
Combined with the considerations presented in subsection $4.1$, this leads to
the identification of $d\eta^i$ with $\delta (\xi^i)$. Note that the scaling
properties of these operators

\eqn\scale{\eta^i \rightarrow \lambda\eta^i \qquad\qquad \delta (\xi^i)
\rightarrow {\delta (\xi^i) \over \lambda}}
imply that $d\eta^1\ldots d\eta^{2m}$ always scales as $W^{-m}$. This
permits a natural conjecture for the holomorphic $\c$-form

\eqn\holom{\Omega = \res [{d\eta^1\ldots d\eta^{2m} \,\omega\over p}],}
where $p=W+\Gamma$. In clear analogy to the Calabi-Yau case, this
construction only makes sense when the super-first Chern class vanishes.
Since the hypersurface is defined by the superconstraint $p=0$, this form
is $d$-closed with $d$ given in \exterior. For these models, there are no
polynomial deformations in the ghost directions. The construction of the
remaining observables is then straightforward. The forms

\eqn\middlecoh{ \varpi (P) = {P(z^1\ldots z^{n+1}) \,d\eta^1\ldots d\eta^{2m}
              \, \omega \over p^k}}
under the residue mapping provide a basis for the middle cohomology of $\sm$.
At least for the single polynomial constraint, these
forms are in a one-to-one correspondence with forms on the body described
in \middle. However, the pole-order is shifted ensuring that these
operators have the correct charge. By adding the forms
$(1,\Omega\wedge{\bar\Omega})$, we recover the Hodge diamond expected
from \lg calculations. Do these forms constitute a complete set for
$\sm$? Physically, we expect no additional forms. However, we offer no
general proof of completeness here. Such a proof requires a suitable
mathematical theory of supercohomology. Nevertheless, we can present some
heuristic expectations. Since $\{ Q, \eta^i \}$ is bosonic, we expect a
closed form to depend on the ghosts only through the constraint $p$.
These are precisely the forms we have just discussed. Any
other closed form  $ \varpi$ - including the \kh form - should be
cohomologically trivial in an appropriate sense.   This would provide a
geometric explanation for the decoupling phenomenon.

The path-integral provides a definition for the integral of the volume
form over the supermanifold,

\eqn\volume{\int_{\sm} {\Omega\wedge{\bar \Omega}}.}
After integrating out the ghost fields $(\eta, \xi)$, the integral
reduces to the body $\cy$ of $\sm$. Expanding $p$ in \holom\ inside the
residue,

\eqn\expansion{{1\over p} = {1\over W} \left\{ 1 - {\Gamma\over W} + \ldots
                \right\},}
and integrating over the $\eta$ fields selects the
term proportional to $W^{-m}$. Integration over the $\xi$ fields is
trivial. After evaluating the residue on the body, we are left with a
integral proportional to the natural volume form on $\cy$. This procedure
agrees with the identification \identify\ if when integrating over $\sm$
we simply perform the Berezin integrals in \holom\ and then evaluate the
residues. This recipe therefore avoids evaluating residues on $\sm$. To show
this procedure agrees with first evaluating the residues and then
computing the path-integral requires a more detailed investigation of
residues than presented here. By comparison with \lg results, we do expect
both procedures to agree, though the proof appears to be quite nontrivial.
Without such a proof, the residue construction remains somewhat formal.
By noting that the form

\eqn\yukawa{\varpi = \res [{q_1\ldots q_{\c} \,d\eta^1\ldots d\eta^{2m}
              \, \omega \over p^{\c+1} }]}
is proportional to ${\bar\Omega}$, the computation of Yukawa couplings is
straightforward. Each $q_i(z^1,\ldots,z^{n+1})$ is a degree $d$
polynomial, and the associated Yukawa coupling is given by:

\eqn\coupling{\kappa (q_1,\ldots ,q_{\c}) = {\int_{\sm} \Omega\wedge\varpi
\over
\int_{\sm} {\Omega\wedge{\bar \Omega}}}.}
After integrating out the ghosts, this expression reduces to intersection
theory on $\cy$. For the case of a single polynomial constraint, this
agrees with the corresponding computation in the \lg theory as previously
discussed \landausigma. However, the constructions described here are
general, and extend to the case of many constraints, more general
embedding spaces etc. These Yukawa couplings then provide a relatively
simple way of computing the instanton corrected couplings of the mirror
theory.

Let us close this discussion by checking the decoupling condition that
the correlator $<k^n \vartheta >$ vanish. This follows from a counting
argument.
The operator
$\vartheta$ must have total charge $2\c - 2n$ by charge conservation.
Clearly $\vartheta$ cannot be $(1,\Omega\wedge{\bar\Omega})$, so
$\vartheta$ is a product of forms each with charge $\c$. Further, each
form with charge $\c$ absorbs $\b$ fermionic $(\psi,{\bar\psi})$ zero modes.
Now $k^n$
absorbs at most $2n$ $(\psi,{\bar\psi})$ zero modes. There are at least
$2\b$ such zero modes. To absorb all the zero modes and satisfy charge
conservation when $n \not= 0 $ requires

\eqn\contradiction{{\b\over\c} \leq 1}
which is a contradiction. Therefore $<k^n \vartheta >$ vanishes and the
\kh form decouples for these models.

\subsec{\kh Moduli}

{}From the previous discussion, we found that the only diagonal Hodge
numbers that were non-zero corresponded to the the identity and volume forms.
How then can \kh moduli arise? The only possibility is from the
resolution of fixed point sets. In the \lg framework, the only models
with \kh moduli correspond to varieties in weighted superprojective
space. Let us take a specific example with superpotential:

\eqn\weightedex{W = (y^1)^6 + (y^2)^6 + (x^1)^3 + \ldots + (x^5)^3 .}
This \lg model has $\c = 3$. The Hodge diamond obtained after orbifolding
by the canonical ${\bf Z}_6$
is shown in figure $4.2$. There are five twisted sectors in the \lg
theory. The untwisted sector provides the forms in the
middle cohomology of the first diamond in figure $4.2$. The identity and
volume form correspond to vacua for the $(1,5)$ twisted sectors \cvlg.
The existence of these forms for the sigma model follows from our
previous discussion. The interesting forms appear in the $(2,3,4)$
twisted sectors and are shown in the second Hodge diamond. These
forms should arise from a resolved fixed-point set.
\vskip 0.1in
$$\matrix{& & & &1 & & & \quad & & \quad & & & &0 & & & \cr
          & & &0 & &0 & & \quad & & \quad & & &0 & &0 & &\cr
          & &0 & &0 & &0 & \quad & & \quad & &0 & &1 & &0 & \cr
          &1 & &68 & &68 & &1 \quad &+ & \quad &0 & &5 & &5 & &0 \cr
          & &0 & &0 & &0 & \quad & & \quad & &0 & &1 & &0 &\cr
          & & &0 & &0 & & \quad & & \quad & & &0 & &0 & &\cr
          & & & &1 & & & \quad & & \quad & & & &0 & & & \cr}$$

\noindent {\ninerm \it Figure 4.2: Hodge diamond for \weightedex\ with
contributions from the $(2,3,4)$ twisted sectors displayed in the second
diamond.}
\vskip 0.1in

The corresponding super-variety $\sm$ has defining constraint

\eqn\moduliex{(y^1)^6 + (y^2)^6 + (x^1)^3 + \ldots + (x^5)^3 + \eta^1
\eta^2 = 0}
with embedding space $\wsp (1,1,2,2,2,2,2
|{\tilde\lambda}_1,{\tilde\lambda}_2)$. The choice of ghost scaling
weights must satisfy ${\tilde\lambda}_1+{\tilde\lambda}_2 = 6$. Ideally,
this theory could be identified with an orbifold of homogeneous
projective space, and studied using the techniques in \zaz. This usually
involves the change of coordinates $x^i = (z^i)^{\lambda_i}$ where $x^i$
has degree $\lambda_i$. Obviously, we cannot make such a change of
variables for the ghost fields. Nevertheless, orbifold considerations
should still be applicable.

The theory should possess a fixed-point set of positive codimension. This
requirement uniquely determines the ghost scaling weights to be
$(2,4)$. A purely bosonic fixed-point set would not provide any
interesting observables for $\sm$. The only fixed point set under the
projective
identification is given by:

\eqn\fixedpt{(x^1)^3 + \ldots + (x^5)^3 + \eta^1 \eta^2 = 0 \qquad\qquad
y^1=y^2=0.}
This $\c=1$ `supercurve' is a ${\bf Z}_2$-quotient singular
set. Forms on $\sm$ which arise from a resolution of this fixed
point set should correspond to observables in the single twisted sector
associated to this ${\bf Z}_2$ quotient. This is clearly heuristic
reasoning but it will prove useful. Observables in the twisted sector
correspond to forms on the fixed-point set \zaz. The cohomology for the
curve is constructed using the techniques of subsection $4.2$; the Hodge
numbers are shown in figure $4.3$.

In the twisted sector, the fermion vacuum is charged which in this case
shifts the charges of the operators in figure $4.3$ by $(1,1)$ \wen. These
observables then provide the missing forms on $\sm$. Note that the forms
$(k,k^2)$ on $\sm$ correspond to
the identity and volume form respectively on the fixed point set.

\vskip 0.1in
$$\matrix{& &1 & \cr
          &5 & &5 \cr
          & &1 & \cr}$$

\noindent {\ninerm \it Figure 4.3: Hodge diamond for the fixed point set.}

\vskip 0.1in\noindent
This line of reasoning, albeit heuristic, implies that the resolved
supermanifold
should possess a \kh modulus. To desingularize $\sm$, we must smoothly
`glue' in an appropriate supermanifold while preserving super-Ricci
flatness. The Hodge diamond for the resulting smooth space should
coincide with figure $4.2$ from the \lg theory.

\vskip 0.4in
\newsec{Conclusions}

To provide a general \lg orbifold with a sigma model phase
requires the introduction of supermanifolds. The condition for conformal
invariance is the vanishing of the super-Ricci tensor. We described the
chiral ring for these theories and argued that \kh moduli only appear when
the target space is singular. Among the supermanifolds considered are
the mirrors of rigid manifolds resolving that issue in mirror symmetry.
Many interesting questions remain to be solved: What are the necessary and
sufficient conditions for the existence of a super-Ricci flat metric? How
do index theorems extend to these spaces? \indexthm\ What is an
appropriate mathematical supercohomology theory? How are singularities
resolved for supermanifolds? Clearly, much of algebraic geometry must
generalize nontrivially to these spaces.

More physically, our intuitive notion of a string propagating on a target
space must be enlarged to accommodate the idea of negative dimensions. Once
again, string theory provides unexpected relations - this time between strings
on bosonic spaces and strings on supermanifolds.  We conclude with a modest
conjecture on mirror symmetry: Any \kh supermanifold giving rise to a
nondegenerate conformal field theory has one or more mirror realizations.

\bigbreak\bigskip\bigskip\centerline{{\bf Acknowledgements}}\nobreak

I would like to thank M. Bershadsky, J. Edwards, A. Klemm, A-K Liu, D.
Morrison, J. Rabin, A. A. Voronov and E. Zaslow for valuable discussions.  I
would especially like to thank C. Vafa for continual guidance and
encouragement. This work was supported in part by a Beinecke Memorial
Fellowship, a Hertz Fellowship, NSF grants PHY-92-18167, PHY-89-57162 and a
Packard Fellowship.

\listrefs
\end